\documentclass[11pt, showpacs]{revtex4}
\usepackage{graphicx, epsfig, axodraw}

\def\beq{\begin{equation}}
\def\eeq{\end{equation}}
\def\beqa{\begin{eqnarray}}
\def\eeqa{\end{eqnarray}}
\def\iar{\begin{array}{l}}
\def\ear{\end{array}}

\begin{document}

\title{Singularity of Feynman propagator and Cutkosky's cutting rules}
\author{Yong Zhou}
\affiliation{China University of Petroleum, College of Physics Science and Technology, 739 Bei'yi Road, Dongying Shandong 257061, China}

\begin{abstract}
We improve on Cutkosky's cutting rules which is used to calculate the contribution of the singularities of Feynman propagators to Feynman amplitude. The correctness of the improved cutting rules is verified by the calculations of the conventional loop momentum integral algorithm. We also find that only the process of the intermediate on shell particles that is same as real physical process can exist in quantum vacuum.
\end{abstract}

\maketitle

\section{Introduction}

Cutkosky's cutting rules is a simple algorithm to calculate the contribution of the singularities of Feynman propagators to Feynman amplitude \cite{c0}, and has been widely used in calculating the imaginary parts of physical amplitudes. Many relevant problems, for example the Abramovskii-Kancheli-Gribov cutting Rules \cite{c1a}, the cutting rules in finite temperature field theory \cite{c1b}, the unitarity method in the cutting rules \cite{c1c}, the Feynman Tree Theorem \cite{c1f}, the dispersion method to calculate Feynman amplitudes \cite{c1g}, the duality relation to calculate Feynman integrals \cite{c1h}, the cutting rules and the unitarity of noncommutative field theory from string theory \cite{c1d}, and the Hopf algebra in the cutting rules \cite{c1e}, have been widely discussed around Cutkosky's cutting rules. From these discussions one finds that Cutkosky's cutting rules is a very useful method both in actual calculations and in searching for underlying physical meanings of Feynman diagrams.

But all of the discussions (including Cutkosky's initial discussion) haven't carefully investigated the algorithm's structure, so haven't noticed the immaturity of the algorithm and the relevant underlying physical meanings. Here we give a careful investigation about it. Consider the Feynman propagator,
\beq
  D_F(x-y)\,=\,\int\frac{d^4 p}{(2\pi)^4}\frac{i}{p^2-m^2+i\varepsilon}\,
  e^{-i p\cdot(x-y)}\,,
\eeq
there are two singularities $p_0=\pm(\sqrt{{\bf p}^2+m^2}-i\varepsilon)$ in the denominator, Since the cutting rules is proposed for calculating the contribution of these singularities to Feynman amplitude, it should point out that in what cases which or both of the singularities have contribution to Feynman amplitude, since in some cases only one of the singularities has contribution to Feynman amplitude, for example in the one loop self energy,
\beqa
  \int\frac{d^4 k}{(2\pi)^4}\frac{-i}{k^2-m_1^2+i\varepsilon}\frac{1}{(k-p)^2-m_2^2+i\varepsilon}\,, \nonumber
\eeqa
where the outline momentum $p$ is on mass shell $p_0=\sqrt{{\bf p}^2+m^2}$ ($m$ is the mass of the outline particle), there are four singularities in the propagators, if keeping all of the contributions of the singularities, we have the result by Cutkosky's cutting rules:
\beqa
  Im{\cal M}(p\rightarrow p)&=&Im\int\frac{d^4 k}{(2\pi)^4}\frac{-i}{k^2-m_1^2+i\varepsilon}
  \frac{1}{(k-p)^2-m_2^2+i\varepsilon} \nonumber \\
  &\rightarrow&\frac{(m^4+m_1^4+m_2^4-2 m^2 m_1^2-2 m^2 m_2^2-2 m_1^2 m_2^2)^{1/2}}
  {16\pi m^2} \nonumber \\
  &\times&\bigl{(} \theta[m_1-m-m_2]+\theta[m-m_1-m_2]+\theta[m_2-m-m_1] \bigr{)}\,,
\eeqa
where $\theta$ is the Heaviside step function, there are three Heaviside step functions (there should be four step functions, but one of them has been eliminated by the energy conservation law in each vertex), obviously such result is wrong, since according to the Breit-Wigner formula $Im {\cal M}(p\rightarrow p)=m\Gamma$ ($\Gamma$ is the decay width of the outline particle) \cite{c2} only the second step function has contribution to the imaginary part of the self energy. There are some discussions about the problem of the selection of the singularities \cite{c1b, c1f, c1g, c1e}, but a thorough and practical prescription hasn't been present.

So setting up an explicit and practicable algorithm for finding the singularities having contribution to Feynman amplitude becomes very important. In order to solve this problem we firstly discuss the origin of the three step functions of Eq.(2). It is obvious from the calculation of Eq.(2) that the three step functions come from the three different combinations of the singularities, as shown in Fig.1.
\vspace{2mm}
\begin{center} \begin{picture}(278,25)
  \SetScale{1.1} \SetWidth{0.45}
  \ArrowLine(0,0)(25,0)
  \ArrowArcn(38,0)(13,180,0)
  \ArrowArcn(38,0)(13,0,180)
  \ArrowLine(50,0)(75,0)
  \Text(10,8)[]{$m$}
  \Text(42,22)[]{$m_1$}
  \Text(42,-23)[]{$m_2$}
  \ArrowLine(85,0)(110,0)
  \ArrowArc(123,0)(13,180,0)
  \ArrowArcn(123,0)(13,180,0)
  \ArrowLine(135,0)(160,0)
  \Text(104,8)[]{$m$}
  \Text(134,22)[]{$m_1$}
  \Text(134,-23)[]{$m_2$}
  \ArrowLine(170,0)(195,0)
  \ArrowArc(208,0)(13,0,180)
  \ArrowArc(208,0)(13,180,0)
  \ArrowLine(220,0)(245,0)
  \Text(198,8)[]{$m$}
  \Text(230,22)[]{$m_1$}
  \Text(230,-23)[]{$m_2$}
\end{picture} \vspace{12mm} \\
{\small FIG. 1: The three different combinations of the singularities of the one loop self energy diagram.}
\end{center}
\vspace{4mm}
In Fig.1 we stipulate that an arrow in a propagator denotes the propagator is on mass shell, and the arrow direction denotes the particle of the propagator incoming into or outgoing from the vertex of the Feynman diagram as convention, and the energy component of the momentum along the arrow direction is positive. Thus an arrow in a propagator represents a singularity of the propagator, and the direction of the arrow tells us which singularity it represents. Since we call a kind of combination of propagators' singularities as a 'cut', the three diagrams in Fig.1 represent three different cuts of the one loop self energy diagram. From the calculation of Eq.(2) we find that the first, the second and the third cuts of Fig.1 contributes the first, the second and the third step functions of Eq.(2) respectively. From the previous discussions we know only the second cut of Fig.1 has contribution to the one loop self energy amplitude, the other two cuts of Fig.1 must be eliminated. Comparing the three cuts of Fig.1 we find the difference between the second and the other cuts is whether the arrow directions are along a contrary circumrotating direction in the loop: in the second cut the arrow directions are along a contrary circumrotating direction in the loop, but in the other two cuts the cases are just opposite. Since in physical meanings the arrow direction stipulated here represents the propagating direction of the intermediate on shell particle in quantum vacuum, the arrow directions being contrary in a loop means the corresponding intermediate on shell particles propagating from a same vertex to another same vertex. If we can generalize this conclusion to the Feynman diagrams with arbitrary number loops and arbitrary number outlines, we will draw the conclusion after some simple analysis that all of the intermediate on shell particles in quantum vacuum propagate along a same direction: from the incoming particles to the outgoing particles of the Feynman diagram.

\section{Improve the cutting rules}

We assume that the above-mentioned generalized conclusion is right, i.e. only the cut in which all of the intermediate on shell
particles propagate along a same direction--from the incoming particles to the outgoing particles of the Feynman diagram--has
contribution to the Feynman amplitude. This means only the cut in which the arrow directions stipulated here of the cut propagators (i.e. the propagating directions of the intermediate on shell particles) cannot form a closed circle in any of the loops has contribution to Feynman amplitude.

In order to suitable for actual calculations we need to solve a potential problem in a kind of Feynman diagrams part of whose innerlines are inserted by self-energy or tadpole diagram. This kind of diagrams, for example see the diagram of Fig.2
\vspace{2mm}
\begin{center} \begin{picture}(82,38)
  \SetWidth{0.40}
  \ArrowLine(0,16)(25,16)
  \CArc(41,16)(16,180,0)
  \CArc(41,16)(16,0,180)
  \GCirc(41,32){6}{0.5}
  \Line(25,16)(57,16)
  \ArrowLine(57,16)(82,16)
  \Text(27,29)[r]{A}
  \Text(56,29)[l]{B}
\end{picture} \vspace{3mm} \\
{\small FIG. 2: A diagram one of whose innerlines is inserted by self-energy diagram.}
\end{center} \vspace{4mm}
In Fig.2 if A and B propagators can be cut simultaneously, it will contribute an infinite large term to the Feynman amplitude when A and B particles have same masses, since there will be two same delta functions in the loop momentum integral. So such cut should be prohibited. In exact words, beyond one cut among the propagators which are connected each other by self-energy or tadpole diagrams is prohibited. Under this constraint the contribution of the cuts of Fig.2 will be equal to $f(m_A^2)/(m_A^2-m_B^2)+f(m_B^2)/(m_B^2-m_A^2)$, where $f$ denotes the amplitude of cutting A propagator and removing the denominator of B propagator or the contrary case, and $m_A, m_B$ is the mass of A and B propagators respectively. When $m_A=m_B=m$, the cut contribution will be equal to $\partial f(m^2)/\partial m^2$. Through some simple deduction we can draw the conclusion that all of the contributions of the cut diagrams which have arbitrary number propagators which are connected each other by self-energy or tadpole diagrams are finite under the above-mentioned constraint. To specific, if the number of the propagators connected each other by self-energy or tadpole diagrams is $n$ and all of the masses of the connected propagators are same, the contribution of the cut diagrams to the Feynman amplitude will be equal to $\partial^n h(m^2)/(\partial m^2)^n$, where $m$ is the same mass of the connected propagators and $h$ is the amplitude of cutting one connected propagator and removing the denominators of all of the other connected propagators.

Now we need to determine the contribution of the singularity of a propagator to Feynman amplitude. Base on the mathematic formula
\beq
  \frac{1}{a\pm i\varepsilon}\,=\,P\frac{1}{a}\mp i\,\pi\,\delta(a)\,,
\eeq
where $P$ denotes the Cauchy principle value, the contribution of a determinate singularity of a propagator, i.e. a cut of the propagator, to Feynman amplitude should be
\beq
  \frac{1}{p^2-m^2+i\varepsilon}\,\rightarrow\,-i\pi\theta[p_0]\,\delta(p^2-m^2)\,,
\eeq
where the propagating direction of the cut propagator has been given, and the term $\theta[p_0]$ is just used to guarantee the energy component of the momentum $p$ along the given propagating direction is positive, as required by the previous stipulation. Since the delta function has two singularities
\beqa
  \delta(p^2-m^2)\,=\,\frac{\delta(p_0-\sqrt{{\bf p}^2+m^2})}{p_0+\sqrt{{\bf p}^2+m^2}} +
  \frac{\delta(p_0+\sqrt{{\bf p}^2+m^2})}{p_0-\sqrt{{\bf p}^2+m^2}}\,, \nonumber
\eeqa
we find that Eq.(4) is different from Cutkosky's formula: there is only one singularity in Eq.(4), but in Cutkosky's formula both of the singularities exist \cite{c0}. We can simplify the expression of Eq.(4) as follow:
\beqa
  \frac{1}{p^2-m^2+i\varepsilon}\,\rightarrow\,-i\pi\delta(p_0-\sqrt{{\bf p}^2+m^2})/
  (2\sqrt{{\bf p}^2+m^2})\,. \nonumber
\eeqa

Furthermore we notice that when we integrate a loop momentum, all of the selected singularities in the loop will be calculated $i$ times where $i$ is the number of the selected singularities in the loop. This is because there are $i$ times integrations for the $i$ number delta functions, so all of the selected singularities will be calculated $i$ times. Therefore if a loop has $i$ number cut propagators, the result should be multiplied by an extra factor $i$. For a complex Feynman diagram the number of loops can be large than one and the loops can be entangled each other, so in order to avoid repeated counting we only need to count the extra factor of the loops which don't contain any smaller loop in it.

In the view of mathematica the principle value and the single delta function in Eq.(3) has similar contribution to Feynman integrals, both contribute normal results, only two or beyond two propagators' delta functions in a momentum loop have abnormal contribution: the discontinuity of the Feynman amplitude. Thus a Feynman amplitude can be divided into three parts: the contribution that all of the Feynman propagators are replaced by their Cauchy principle values, the contribution that part of the Feynman propagators are cut (i.e. replaced by the delta functions) but none of the momentum loops are cut beyond once, and the contribution that there are momentum loops which are cut beyond once. Obviously only the last part contributes the discontinuity to the Feynman amplitude. Since the purpose of proposing the cutting rules is to calculate the discontinuity of any Feynman diagram \cite{c0, c2b}, the algorithm should reach such goal: all of the discontinuity part of a Feynman diagram are denoted by cut, and the remaining part aren't denoted by cut. According to such requirement the cutting rules should satisfy the following two conditions: 1) the least number of the cut propagators in a cut momentum loop is two; 2) the uncut propagators in a cut momentum loop are replaced by their Cauchy principle values, the propagators in an uncut momentum loop keep unchanged, but their contribution to the discontinuity of the Feynman diagram needs to be removed.

Summing up all of the above-mentioned discussions we get the following improved cutting rules:
\begin{enumerate}
\item Cut through the Feynman diagram in all possible ways such that the cut propagators can simultaneously be put on mass shell and the cut propagators' propagating directions (it is required that the energy component of the momentum along the propagating direction of the cut propagator is positive) cannot form a closed circle in any loop; keep the number of the cut propagators in a same propagator chain (where the propagators in a same propagator chain means all of the propagators are connected each other by self-energy or tadpole diagrams) less than two; keep the least number of the cut propagators in any cut loop is two.
\item For each cut propagator with definite propagating direction, replace $1/(p^2-m^2+i\varepsilon)\rightarrow -i\pi\delta(p_0-\sqrt{{\bf p}^2+m^2})/(2\sqrt{{\bf p}^2+m^2})$ where the momentum $p=(p_0,{\bf p})$ is along the given propagating direction of the cut propagator; for each uncut propagator in a cut loop, apply Cauchy principle value to it; for the propagators in an uncut loop, keep them unchanged, but their contribution to the discontinuity of the Feynman diagram needs to be removed; for each cut loop which doesn't contain any smaller loop in it, multiply by factor $i$ where $i$ is the number of the cut propagators in the loop, then perform the loop integrals.
\item Sum the contributions of all possible cuts.
\end{enumerate}

We give an example to illustrate the improved cutting rules. Fig.3 shows the cuts contributing real part to the two loop self energy according to the improved cutting rules. From Fig.3 we can see several different processes happen in the quantum vacuum that the incoming particle decays into some intermediate on shell particles then these intermediate on shell particles decay into the outgoing particle. In addition we find that the singularities of Feynman propagators also contribute real part to Feynman amplitude.
\setcounter{figure}{2}
\begin{figure}[htbp]
\begin{center}
  \epsfig{file=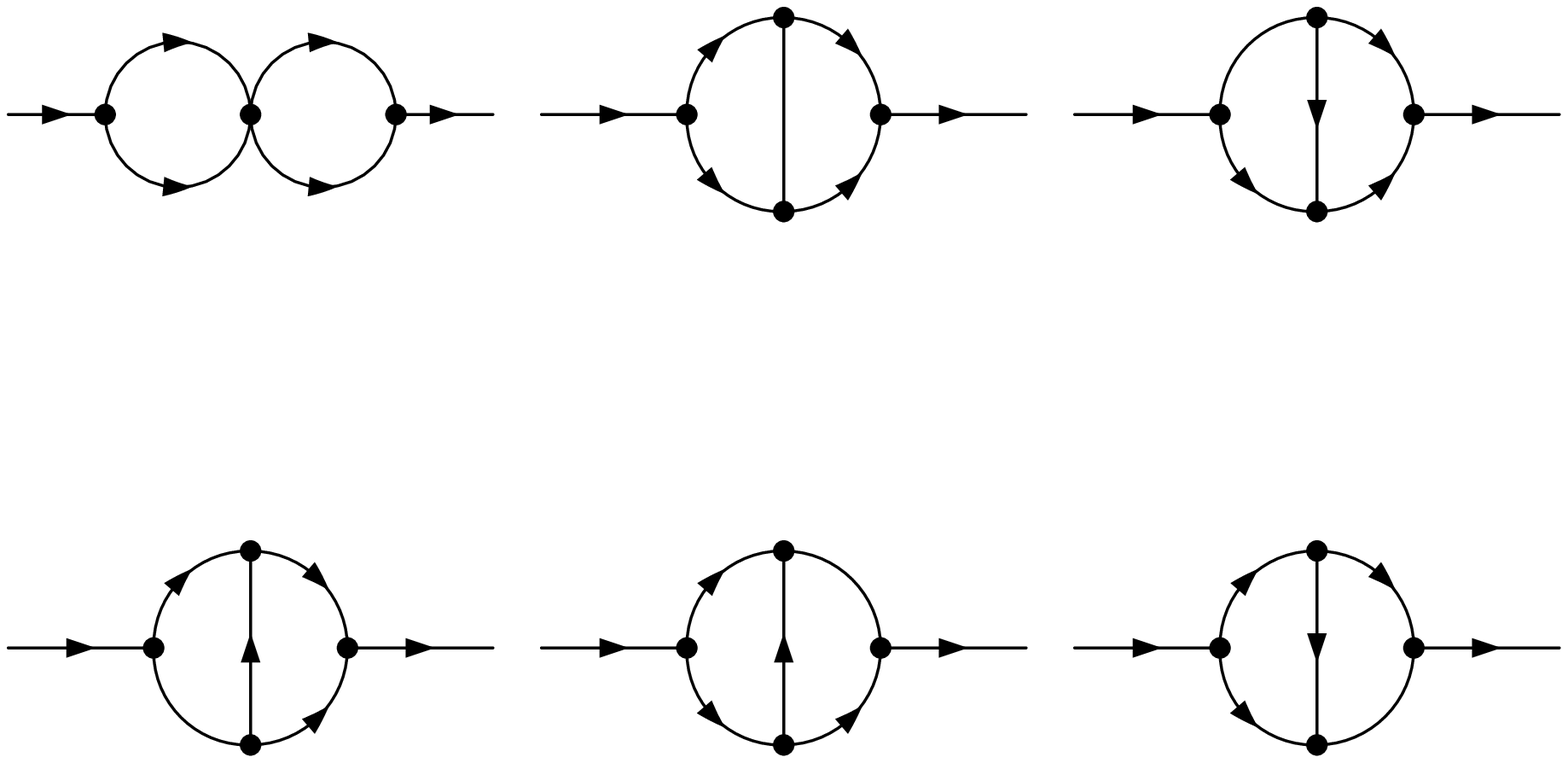,width=8.7cm} \\
  \caption{Cuts contributing real part to the two loop self energy.}
\end{center}
\end{figure}

\section{Correctness of the improved cutting rules}

In this section we will examine the correctness of the improved cutting rules. We will calculate the imaginary parts of some Feynman diagrams by two methods. One method is the improved cutting rules, the other method is the conventional integral algorithm, i.e. the Feynman parametrization, wick rotation and the dimensional regularization \cite{c4}. We will compare the two methods' results to judge whether the improved cutting rules is correct. In the following calculations we have used the program packages {\em FeynArts} and {\em FeynCalc} \cite{c6}. The examples of calculating the contributions of the singularities to real parts of Feynman amplitudes can be found in Ref.\cite{c5}.

Firstly we calculate the imaginary part of the two-loop two-point Feynman diagram shown in Fig.4, where the cut has been drawn and $p^2=m^2$.
\vspace{5mm}
\begin{center} \begin{picture}(80,20)
\SetScale{1.35} \SetWidth{0.30}
  \ArrowLine(0,2)(25,2)
  \ArrowLine(25,2)(50,2)
  \ArrowLine(50,2)(75,2)
  \ArrowArcn(38,2)(13,180,0)
  \ArrowArc(38,2)(13,180,0)
  \Text(14,10)[]{\small $p$}
  \Text(52,26)[]{\small $m_1$}
  \Text(52,10)[]{\small $m_1$}
  \Text(52,-8)[]{\small $m_1$}
  \Text(83,10)[]{$m$}
\end{picture} \vspace{9mm} \\
{\small FIG. 4: Cut contributing imaginary part of a two-loop two-point diagram.}
\end{center}
\vspace{4mm}
Using the improved cutting rules we obtain
\beqa
  &&Im\int\frac{d^4 k_1}{(2\pi)^4}\frac{d^4 k_2}
  {(2\pi)^4}\frac{1}{k_1^2-m_1^2+i\varepsilon}\frac{1}{k_2^2-m_1^2+i\varepsilon}\frac{1}
  {(k_1+k_2-p)^2-m_1^2+i\varepsilon} \nonumber \\
  =\hspace{-3mm}&&\frac{1}{64\pi^5}\int d^4 k_1 d^4 k_2\,\theta[k_{10}]\,
  \delta(k_1^2-m_1^2)\,\theta[k_{20}]\,\delta(k_2^2-m_1^2)\,\theta[p_0-k_{10}-k_{20}]\,
  \delta((p-k_1-k_2)^2-m_1^2) \nonumber \\
  =\hspace{-3mm}&&\frac{1}{64 m\pi^3}\int^{\frac{m^2-3 m_1^2}{2 m}}_{m_1} d\,x
  \Bigl{[}(m x+m_1^2)^2-\frac{m_1^2(m^2-m_1^2)^2}{m^2+m_1^2-2 m x}\Bigr{]}^{1/2}
  \theta[m-3m_1]\,.
\eeqa
Using the conventional integral algorithm we also obtain
\beqa
  &&Im\int\frac{d^D k_1}{(2\pi)^D}
  \frac{d^D k_2}{(2\pi)^D}\frac{1}{k_1^2-m_1^2+i\varepsilon}
  \frac{1}{k_2^2-m_1^2+i\varepsilon}\frac{1}{(k_1+k_2-p)^2-m_1^2+i\varepsilon}
  \nonumber \\
  =\hspace{-3mm}&&\frac{1}{256\pi^4}Im\int_0^1 d\,x\int_0^1 d\,y\,
  (A-2 m^2 y)(\frac{2}{\epsilon}-2\gamma+\ln 16\pi^2+\ln y-\ln x(1-x)-\ln B)\ln B
  \nonumber \\
  =\hspace{-3mm}&&-\frac{1}{256\pi^3}\int_0^1 d\,x\int_0^1 d\,y\,
  (A-2 m^2 y)(\frac{2}{\epsilon}-2\gamma+\ln 16\pi^2+\ln y-\ln x(1-x)-2\ln|B|)
  \theta[-B] \nonumber \\
  =\hspace{-3mm}&&\frac{1}{128\pi^3}\int_{x_1}^{x_2}d\,x\Bigl{(}
  \frac{A(A^2-4 m^2 m_1^2)^{1/2}}{4 m^2}-m_1^2\ln\frac{A+(A^2-4 m^2 m_1^2)^{1/2}}
  {2 m m_1}\Bigr{)}\theta[m-3 m_1]\,,
\eeqa
where $\epsilon=4-D$, $\gamma$ is the Euler constant, and
\beqa
  A\,=\hspace{-3mm}&&m^2+m_1^2-\frac{m_1^2}{x(1-x)}\,, \nonumber \\
  B\,=\hspace{-3mm}&&m^2 y^2-A\,y+m_1^2\,, \nonumber \\
  x_{1,2}\,=\hspace{-3mm}&&\frac{m-m_1\mp(m^2-2 m m_1-3 m_1^2)^{1/2}}{2(m-m_1)}\,.
\eeqa
In Eq.(6) we have used the formula $\ln a=\ln|a|-i\pi\,\theta[-a]$. Both Eq.(5) and Eq.(6) are too complex to be expressed by analytical formula, so we only compare their numerical results as shown in Fig.5. From Fig.5 we find that the results obtained by the two methods coincide very well.
\setcounter{figure}{4}
\begin{figure}[htbp]
\begin{center}
  \epsfig{file=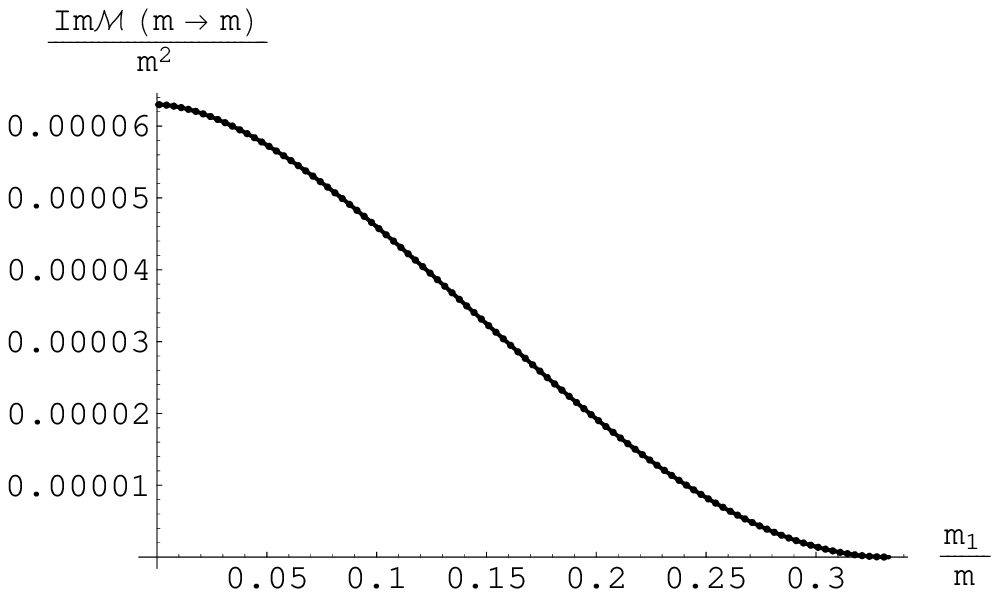, width=7cm} \\
  \caption{Numerical results of Eq.(5) (denoted by dots) and Eq.(6) (denoted by line).}
\end{center}
\end{figure}
In order to show to what extent the coincidence is we calculate the relative discrepancy of the two results, as shown in Fig.6.
\begin{figure}[htbp]
\begin{center}
  \epsfig{file=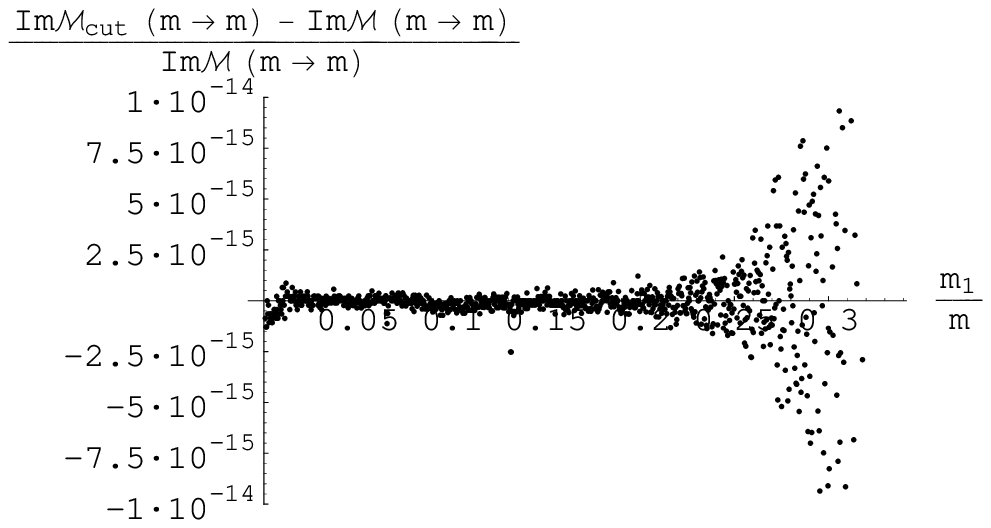, width=7cm} \\
  \caption{The relative discrepancy of Eq.(5) and Eq.(6). The amplitude with a subscript "cut" denotes
  it is obtained by the improved cutting rules, and the amplitude without subscript is obtained
  by the conventional integral algorithm.}
\end{center}
\end{figure}
We note that we have used the software {\em Mathematica 5.0} to calculate the result, and we have set the precision goal as $10^{-10}$. The maximum relative discrepancy of Eq.(5) and Eq.(6) is $3.4\times 10^{-11}$ which is less than the set precision goal, so we can draw the conclusion that Eq.(5) is equal to Eq.(6) at the precision $10^{-10}$. Besides, we haven't shown the data whose absolute values are greater than $10^{-14}$ in Fig.6 in order to make the great majority of the data clearly shown in the graph.

Then we calculate the imaginary part of the one-loop three-point Feynman diagram shown in Fig.7, where $p^2=m_a^2$ and $p_1^2=p_2^2=m_b^2$.
\vspace{4mm}
\begin{center} \begin{picture}(335,25)
  \SetScale{1.1} \SetWidth{0.45}
  \ArrowLine(0,0)(25,0)
  \Line(25,0)(46,13)
  \Line(25,0)(46,-13)
  \ArrowLine(46,13)(60,20)
  \ArrowLine(46,-13)(60,-20)
  \Line(46,13)(46,-13)
  \Text(10,6)[]{$p$}
  \Text(31,12)[]{$m_1$}
  \Text(31,-12)[]{$m_1$}
  \Text(60,0)[]{$m_1$}
  \Text(74,20)[]{$p_1$}
  \Text(74,-20)[]{$p_2$}
  \Text(89,0)[]{$\bf \longrightarrow$}
  \SetOffset(103,0)
  \ArrowLine(0,0)(25,0)
  \ArrowLine(25,0)(46,13)
  \ArrowLine(46,13)(60,20)
  \ArrowLine(25,0)(46,-13)
  \ArrowLine(46,-13)(60,-20)
  \Line(46,13)(46,-13)
  \Text(77,0)[]{$+$}
  \SetOffset(113,0)
  \ArrowLine(70,0)(95,0)
  \ArrowLine(95,0)(116,13)
  \ArrowLine(116,13)(130,20)
  \Line(95,0)(116,-13)
  \ArrowLine(116,-13)(130,-20)
  \ArrowLine(116,-13)(116,13)
  \Text(154,0)[]{$+$}
  \SetOffset(123,0)
  \ArrowLine(140,0)(165,0)
  \Line(165,0)(186,13)
  \ArrowLine(186,13)(200,20)
  \ArrowLine(165,0)(186,-13)
  \ArrowLine(186,-13)(200,-20)
  \ArrowLine(186,13)(186,-13)
\end{picture} \vspace{12mm} \\
{\small FIG. 7: Cuts contributing imaginary part of a one-loop three-point diagram.}
\end{center}
\vspace{4mm}
Using the improved cutting rules we obtain
\beqa
  &&Im(-i)\int\frac{d^4 k}{(2\pi)^4}\frac{1}{k^2-m_1^2+i\varepsilon}
  \frac{1}{(k-p_1)^2-m_1^2+i\varepsilon}\frac{1}{(k-p_1-p_2)^2-m_1^2+i\varepsilon}
  \nonumber \\
  =\hspace{-3mm}&&P\int\frac{d^4 k}{8\pi^2}\biggl{[} \frac{\theta[k_0]\delta(k^2-m_1^2)
  \theta[p_{10}+p_{20}-k_0]\delta((p_1+p_2-k)^2-m_1^2)}{(k-p_1)^2-m_1^2+i\varepsilon}
  \nonumber \\
  +\hspace{-3mm}&&\frac{\theta[k_0]\delta(k^2-m_1^2)\theta[p_{10}-k_0]
  \delta((p_1-k)^2-m_1^2)}{(k-p_1-p_2)^2-m_1^2+i\varepsilon} \nonumber \\
  +\hspace{-3mm}&&\frac{\theta[k_0-p_{10}]\delta((k-p_1)^2-m_1^2)\theta[p_{10}+p_{20}-k_0]
  \delta((p_1+p_2-k)^2-m_1^2)}{k^2-m_1^2+i\varepsilon} \biggr{]} \nonumber \\
  =\hspace{-3mm}&&\frac{\theta[m_a-2 m_1]}{16\pi m_a\sqrt{m_a^2-4 m_b^2}}
  \ln\left|\frac{m_a^2-2m_b^2-\sqrt{(m_a^2-4 m_b^2)(m_a^2-4 m_1^2)}}{m_a^2-2m_b^2+
  \sqrt{(m_a^2-4 m_b^2)(m_a^2-4 m_1^2)}}\right| \nonumber \\
  +\hspace{-3mm}&&\frac{\theta[m_b-2m_1]}{8\pi m_a\sqrt{m_a^2-4 m_b^2}}
  \ln\left|\frac{m_a m_b+\sqrt{(m_a^2-4 m_b^2)(m_b^2-4 m_1^2)}}
  {m_a m_b-\sqrt{(m_a^2-4 m_b^2)(m_b^2-4 m_1^2)}}\right|\,.
\eeqa
Using the conventional integral algorithm we also obtain
\beqa
  &&Im(-i)\int\frac{d^4 k}{(2\pi)^4}\frac{1}{k^2-m_1^2+i\varepsilon}
  \frac{1}{(k-p_1)^2-m_1^2+i\varepsilon}\frac{1}{(k-p_1-p_2)^2-m_1^2+i\varepsilon}
  \nonumber \\
  &=&-\frac{1}{16\pi^2}\,Im\int_0^1 d\,x\int_0^{1-x} \frac{d\,y}
  {m_a^2\,x^2+m_b^2\,y^2+m_a^2\,x\,y-m_a^2\,x-m_b^2\,y+m_1^2-i\varepsilon} \nonumber \\
  &=&-\frac{1}{16\pi}\int_0^1 d\,x\int_0^{1-x}d\,y\,
  \delta(m_a^2\,x^2+m_b^2\,y^2+m_a^2\,x\,y-m_a^2\,x-m_b^2\,y+m_1^2) \nonumber \\
  &=&\frac{1}{16\pi m_a\sqrt{m_a^2-4 m_b^2}}\ln \biggl{[}
  \frac{m_a^2-2 m_b^2-\sqrt{(m_a^2-4 m_b^2)(m_a^2-4 m_1^2)}}
  {m_a^2-2 m_b^2+\sqrt{(m_a^2-4 m_b^2)(m_a^2-4 m_1^2)}} \nonumber \\
  &\times&\left(\frac{m_a m_b+\sqrt{(m_a^2-4 m_b^2)(m_b^2-4 m_1^2)}}
  {m_a m_b-\sqrt{(m_a^2-4 m_b^2)(m_b^2-4 m_1^2)}} \right)^2 \biggr{]}
  \theta[\frac{m_b}{2}-m_1]+\frac{1}{16\pi m_a\sqrt{m_a^2-4 m_b^2}} \nonumber \\
  &\times&\ln\frac{m_a^2-2 m_b^2-\sqrt{(m_a^2-4 m_b^2)(m_a^2-4 m_1^2)}}
  {m_a^2-2 m_b^2+\sqrt{(m_a^2-4 m_b^2)(m_a^2-4 m_1^2)}}\,
  \theta[\frac{m_a}{2}-m_1]\,\theta[m_1-\frac{m_b}{2}]\,.
\eeqa
In Eq.(9) we have used the formula Eq.(3). Obviously Eq.(8) is equal to Eq.(9). Eq.(8) also coincides with Eq.(3.24) of Ref.\cite{c3}.

Thirdly we calculate the imaginary part of the one-loop four-point Feynman diagram shown in Fig.8, where $r=(2 p, p\,{\bf e_z})$ and $q=(2 p, -p\,{\bf e_z})$.
\vspace{5mm}
\begin{center} \begin{picture}(380,90)
  \SetOffset(0,55)
  \ArrowLine(0,0)(25,0)
  \Line(25,0)(50,0)
  \ArrowLine(50,0)(75,0)
  \Line(25,0)(25,25)
  \Line(50,0)(50,25)
  \ArrowLine(0,25)(25,25)
  \Line(25,25)(50,25)
  \ArrowLine(50,25)(75,25)
  \Text(38,31)[]{$m$}
  \Text(38,-6)[]{$m$}
  \Text(18,13)[]{$m$}
  \Text(57,13)[]{$0$}
  \Text(-6,25)[]{$r$}
  \Text(-6,0)[]{$q$}
  \Text(81,25)[l]{$r$}
  \Text(81,0)[l]{$q$}
  \Text(100,12.5)[]{$\longrightarrow$}
  \SetOffset(90,0)
  \ArrowLine(25,55)(50,55)
  \ArrowLine(50,55)(75,55)
  \ArrowLine(75,55)(100,55)
  \Line(50,80)(50,55)
  \Line(75,55)(75,80)
  \ArrowLine(25,80)(50,80)
  \ArrowLine(50,80)(75,80)
  \ArrowLine(75,80)(100,80)
  \Text(110,68)[]{$+$}
  \SetOffset(210,55)
  \ArrowLine(0,0)(25,0)
  \Line(25,0)(50,0)
  \ArrowLine(50,0)(75,0)
  \ArrowLine(25,25)(25,0)
  \Line(50,0)(50,25)
  \ArrowLine(0,25)(25,25)
  \ArrowLine(25,25)(50,25)
  \ArrowLine(50,25)(75,25)
  \SetOffset(220,55)
  \Text(75,12.5)[]{$+$}
  \ArrowLine(85,0)(110,0)
  \ArrowLine(110,0)(135,0)
  \ArrowLine(135,0)(160,0)
  \ArrowLine(110,0)(110,25)
  \Line(135,0)(135,25)
  \ArrowLine(85,25)(110,25)
  \Line(110,25)(135,25)
  \ArrowLine(135,25)(160,25)
  \SetOffset(-55,0)
  \Text(155,12.5)[]{$+$}
  \ArrowLine(170,0)(195,0)
  \Line(195,0)(220,0)
  \ArrowLine(220,0)(245,0)
  \Line(195,25)(195,0)
  \ArrowLine(220,0)(220,25)
  \ArrowLine(170,25)(195,25)
  \ArrowLine(195,25)(220,25)
  \ArrowLine(220,25)(245,25)
  \Text(255,12.5)[]{$+$}
  \SetOffset(-45,0)
  \ArrowLine(255,0)(280,0)
  \ArrowLine(280,0)(305,0)
  \ArrowLine(305,0)(330,0)
  \Line(280,25)(280,0)
  \ArrowLine(305,25)(305,0)
  \ArrowLine(255,25)(280,25)
  \Line(280,25)(305,25)
  \ArrowLine(305,25)(330,25)
\end{picture} \vspace{6mm} \\
{\small FIG. 8: Cuts contributing imaginary part of a one-loop four-point diagram.}
\end{center} \vspace{4mm}
Using the improved cutting rules we obtain
\beqa
  &&Im(-i)\int\frac{d^4 k}{(2\pi)^4}\frac{1}{k^2+i\varepsilon}
  \frac{1}{(k-q)^2-m^2+i\varepsilon}\frac{1}{k^2-m^2+i\varepsilon}
  \frac{1}{(k+r)^2-m^2+i\varepsilon} \nonumber \\
  =\hspace{-3mm}&&\frac{1}{8\pi^2}P\int d^4 k \Bigl{[} \frac{\theta[k_0]\delta(k^2-m^2)
  \theta[q_0+r_0-k_0]\delta((q+r-k)^2-m^2)}{(k-r)^2((k-r)^2-m^2)} \nonumber \\
  +\hspace{-3mm}&&\frac{\theta[k_0]\delta(k^2-m^2)\theta[r_0-k_0]\delta((r-k)^2-m^2)}
  {k^2((k+q)^2-m^2)}+\frac{\theta[k_0]\delta(k^2-m^2)\theta[q_0-k_0]\delta((q-k)^2-m^2)}
  {k^2((k+r)^2-m^2)} \nonumber \\
  +\hspace{-3mm}&&\frac{\theta[k_0]\delta(k^2)\theta[r_0-k_0]\delta((r-k)^2-m^2)}
  {(k^2-m^2)((k+q)^2-m^2)}+\frac{\theta[k_0]\delta(k^2)\theta[q_0-k_0]\delta((q-k)^2-m^2)}
  {(k^2-m^2)((k+r)^2-m^2)} \Bigr{]} \nonumber \\
  =\hspace{-3mm}&&-\frac{\theta[p-m/\sqrt{3}]}{64\pi m^2 p^2}\ln3+
  \frac{\theta[p-m/2]}{128\pi m^2 p^2}\ln\frac{9p^3-m^2 p+2m^2
  \sqrt{4p^2-m^2}}{9p^3-m^2 p-2m^2\sqrt{4p^2-m^2}} \nonumber \\
  +\hspace{-3mm}&&\frac{\theta[p-2m/\sqrt{3}]}{128\pi m^2 p^2}\ln
  \frac{15p^2-4m^2+4p\sqrt{9p^2-12m^2}}{15p^2-4m^2-4p\sqrt{9p^2-12m^2}}\,.
\eeqa
Using the conventional integral algorithm and Eq.(3) we also obtain
\beqa
  &&Im(-i)\int\frac{d^4 k}{(2\pi)^4}\frac{1}{k^2+i\varepsilon}
  \frac{1}{(k-q)^2-m^2+i\varepsilon}\frac{1}{k^2-m^2+i\varepsilon}
  \frac{1}{(k+r)^2-m^2+i\varepsilon} \nonumber \\
  =\hspace{-3mm}&&\frac{1}{16\pi^2}\,Im\int_0^1 d\,x\int_0^{1-x}d\,y\int_0^{1-x-y}
  \frac{d\,z}{(3 p^2\,x^2+3 p^2\,y^2-10 p^2\,x\,y-3 p^2\,x-3 p^2\,y-m^2\,z+m^2
  -i\varepsilon)^2} \nonumber \\
  =\hspace{-3mm}&&\frac{1}{16\pi^2 m^2}\,Im\int_0^1 d\,x\int_0^{1-x}d\,y\,\left[
  \frac{1}{3 p^2\,x^2+3 p^2\,y^2-10 p^2\,x\,y+(m^2-3 p^2)x+(m^2-3 p^2)y-i\varepsilon}
  \right. \nonumber \\
  -\hspace{-3mm}&&\frac{1}{3 p^2\,x^2+3 p^2\,y^2-10 p^2\,x\,y-3 p^2\,x-3 p^2\,y+m^2
  -i\varepsilon} \biggr{]} \nonumber \\
  =\hspace{-3mm}&&\frac{1}{16\pi m^2}\int_0^1 d\,x\int_0^{1-x}d\,y\,\left[
  \delta(3 p^2\,x^2+3 p^2\,y^2-10 p^2\,x\,y+(m^2-3 p^2)x+(m^2-3 p^2)y)
  \right. \nonumber \\
  -\hspace{-3mm}&&\delta(3 p^2\,x^2+3 p^2\,y^2-10 p^2\,x\,y-3 p^2\,x-3 p^2\,y+m^2)
  \bigr{]} \nonumber \\
  =\hspace{-3mm}&&\frac{\theta[p-\frac{m}{2}]\theta[\frac{m}{\sqrt{3}}-p]}{128\pi m^2 p^2}
  \ln\frac{9p^3-m^2 p+2m^2\sqrt{4p^2-m^2}}{9p^3-m^2 p-2m^2\sqrt{4p^2-m^2}} \nonumber \\
  +\hspace{-3mm}&&\frac{\theta[p-\frac{m}{\sqrt{3}}]\theta[\frac{2m}{\sqrt{3}}-p]}
  {128\pi m^2 p^2}\ln\frac{9p^3-m^2 p+2m^2\sqrt{4p^2-m^2}}{9(9p^3-m^2 p-2m^2
  \sqrt{4p^2-m^2})} \nonumber \\
  +\hspace{-3mm}&&\frac{\theta[p-\frac{2m}{\sqrt{3}}]}{128\pi m^2 p^2}
 \ln\frac{(9p^3-m^2 p+2m^2\sqrt{4p^2-m^2})(15p^2+4p\sqrt{9p^2-12m^2}-4m^2)}
  {9(9p^3-m^2 p-2m^2\sqrt{4p^2-m^2})(15p^2-4p\sqrt{9p^2-12m^2}-4m^2)} \,.
\eeqa
Obviously Eq.(10) is equal to Eq.(11).

Lastly we calculate the imaginary part of the two-loop two-point Feynman diagram shown in Fig.9, where $p^2=M^2$.
\vspace{5mm}
\begin{center} \begin{picture}(415,36)
  \SetScale{0.85} \SetWidth{0.55}
  \ArrowLine(0,0)(30,0)
  \ArrowLine(70,0)(100,0)
  \CArc(50,0)(20,0,65)
  \BCirc(50,20){9}
  \CArc(50,0)(20,115,180)
  \CArc(50,0)(20,180,0)
  \Text(10,8)[]{$p$}
  \Text(26,14)[]{0}
  \Text(46,32)[]{$m$}
  \Text(46,5)[]{$m$}
  \Text(46,-25)[]{$m$}
  \Text(64,14)[]{$m$}
  \Text(105,0)[]{$\longrightarrow$}
  \SetOffset(120,0)
  \ArrowLine(0,0)(30,0)
  \ArrowLine(70,0)(100,0)
  \CArc(50,0)(20,0,65)
  \ArrowArcn(50,20)(9,180,0)
  \ArrowArc(50,20)(9,180,0)
  \CArc(50,0)(20,115,180)
  \ArrowArc(50,0)(20,180,0)
  \Text(100,0)[]{$+$}
  \SetOffset(110,0)
  \ArrowLine(130,0)(160,0)
  \ArrowLine(200,0)(230,0)
  \ArrowArcn(180,0)(20,180,115)
  \BCirc(180,20){9}
  \CArc(180,0)(20,0,65)
  \ArrowArc(180,0)(20,180,0)
  \Text(217,0)[]{$+$}
  \SetOffset(100,0)
  \ArrowLine(260,0)(290,0)
  \ArrowLine(330,0)(360,0)
  \CArc(310,0)(20,115,180)
  \BCirc(310,20){9}
  \ArrowArcn(310,0)(20,65,0)
  \ArrowArc(310,0)(20,180,0)
\end{picture} \vspace{12mm} \\
{\small FIG. 9: Cuts contributing imaginary part of a two-loop two-point diagram.}
\end{center}
\vspace{4mm}
Using the improved cutting rules we obtain
\beqa
  &&Im\,\mu^{2\epsilon}\int\frac{d^D k_1}{(2\pi)^D}\frac{d^D k_2}{(2\pi)^D}
  \frac{1}{k_1^2+i\varepsilon}\frac{1}{k_2^2-m^2+i\varepsilon} \nonumber \\
  \times\hspace{-3mm}&&\frac{1}{(k_2-k_1)^2-m^2+i\varepsilon}\frac{1}{k_1^2-m^2+i\varepsilon}
  \frac{1}{(k_1-p)^2-m^2+i\varepsilon} \nonumber \\
  = \hspace{-3mm} &&Im\,\mu^{2\epsilon}P\int\frac{d^D k_1}{(2\pi)^D}\frac{d^D k_2}{(2\pi)^D}
  \Bigl{[} \frac{1}{(p-k_1)^2((p-k_1)^2-m^2)} \nonumber \\
  \times\hspace{-3mm}&&4i\,\pi^3\,\theta[k_{10}]\delta(k_1^2-m^2)\theta[k_{20}]\delta(k_2^2-m^2)
  \theta[p_0-k_{10}-k_{20}]\delta((p-k_1-k_2)^2-m^2) \nonumber \\
  -\hspace{-3mm}&&\frac{2\pi^2\theta[k_{10}]\delta(k_1^2)\theta[p_0-k_{10}]
  \delta((p-k_1)^2-m^2)}{(k_2^2-m^2)((k_2-k_1)^2-m^2)(k_1^2-m^2)} \nonumber \\
  -\hspace{-3mm}&&\frac{2\pi^2\theta[k_{10}]\delta(k_1^2-m^2)\theta[p_0-k_{10}]
  \delta((p-k_1)^2-m^2)}
  {k_1^2(k_2^2-m^2)((k_2-k_1)^2-m^2)} \Bigr{]} \nonumber \\
  =\hspace{-3mm}&&\frac{1}{64\pi^3}P\int_0^{\infty}k_1^2\,d\,k_1
  \int_0^{\infty}k_2^2\,d\,k_2\int_{-1}^1 d\,x \nonumber \\
  \times\hspace{-3mm}&&\frac{\delta(M-\sqrt{m^2+k_1^2}-
  \sqrt{m^2+k_2^2}-\sqrt{m^2+k_1^2+k_2^2+2 k_1 k_2\,x})}{\sqrt{m^2+k_1^2}
  \sqrt{m^2+k_2^2}\sqrt{m^2+k_1^2+k_2^2+2 k_1 k_2\,x}} \nonumber \\
  \times\hspace{-3mm}&&\frac{1}{(M^2+m^2-2M\sqrt{m^2+k_1^2})(M^2-2M\sqrt{m^2+k_1^2})}
  \nonumber \\
  +\hspace{-3mm}&&\frac{2\pi^2\mu^{2\epsilon}\Gamma(\epsilon/2)}{(4\pi)^{D/2}m^2}\int
  \frac{d^D k_1}{(2\pi)^D}\theta[k_{10}]\delta(k_1^2)\theta[p_0-k_{10}]
  \delta((p-k_1)^2-m^2) \nonumber \\
  -\hspace{-3mm}&&\frac{2\pi^2\mu^{2\epsilon}\Gamma(\epsilon/2)}{(4\pi)^{D/2}m^2}\int_0^1
  \frac{d\,x}{(x^2-x+1)^{\epsilon/2}}\int\frac{d^D k_1}{(2\pi)^D}
  \theta[k_{10}]\delta(k_1^2-m^2)\theta[p_0-k_{10}]\delta((p-k_1)^2-m^2) \nonumber \\
  =\hspace{-3mm}&&\frac{\theta[M-3m]}{64\pi^3 M}m^2\int_1^{\frac{M^3-3m^2}{2m\,M}}
  \frac{\sqrt{(x^2-1)(M^2-2m\,M\,x-3m^2)}}{(M^2+m^2-2m\,M\,x)^{3/2}(M-2m\,x)}
  d\,x \nonumber \\
  +\hspace{-3mm}&&\frac{\theta[M-m]}{256\pi^3 m^2 M^2}(M^2-m^2)
  (\Delta-2\ln\frac{M^2-m^2}{m\,M}+2) \nonumber \\
  -\hspace{-3mm}&&\frac{\theta[M-2m]}{256\pi^3 m^2 M}\sqrt{M^2-4m^2}
  (\Delta-\ln\frac{M^2-4m^2}{m^2}+4-\frac{\pi}{\sqrt{3}})\,,
\eeqa
where
\beq
  \Delta\,=\,\frac{2}{\epsilon}+2\ln4\pi-2\gamma-2\ln\frac{m^2}{\mu^2}\,.
\eeq
Using the conventional integral algorithm and Eq.(3) and the formula $\ln a=\ln|a|-i\pi\theta[-a]$ we also obtain
\beqa
  &&Im\,\mu^{2\epsilon}\int\frac{d^D k_1}{(2\pi)^D}\frac{d^D k_2}{(2\pi)^D}
  \frac{1}{k_1^2+i\varepsilon}\frac{1}{k_2^2-m^2+i\varepsilon} \nonumber \\
  \times\hspace{-3mm}&&\frac{1}{(k_2-k_1)^2-m^2+i\varepsilon}\frac{1}{k_1^2-m^2+i\varepsilon}
  \frac{1}{(k_1-p)^2-m^2+i\varepsilon} \nonumber \\
  =\hspace{-3mm}&&\frac{1}{256\pi^4}Im\int_0^1 d\,x\int_0^1 d\,y\int_0^{1-y}
  \frac{d\,z}{M^2 y^2+(m^2-M^2)y+m^2 z-i\varepsilon}\Bigl{[} \Delta+
  \ln\frac{1-y-z}{x-x^2} \nonumber \\
  -\hspace{-3mm}&&\ln(\frac{M^2 y^2}{m^2}-\frac{M^2 y}{m^2}+y+z
  +\frac{1-y-z}{x-x^2}-i\varepsilon)-\ln(\frac{M^2 y^2}{m^2}-\frac{M^2 y}{m^2}+y+z
  -i\varepsilon) \Bigr{]} \nonumber \\
  =\hspace{-3mm}&&\frac{1}{256\pi^3}\int_0^1 d\,x\int_0^1 d\,y\int_0^{1-y}d\,z\,
  \delta(M^2 y^2+(m^2-M^2)y+m^2 z)\Bigl{[} \Delta+\ln\frac{1-y-z}{x-x^2} \nonumber \\
  -\hspace{-3mm}&&\ln|\frac{M^2 y^2}{m^2}-\frac{M^2 y}{m^2}+y+z
  +\frac{1-y-z}{x-x^2}| \Bigr{]} \nonumber \\
  +\hspace{-3mm}&&\frac{1}{256\pi^3}P\int_0^1 d\,x\int_0^1 d\,y\int_0^{1-y}d\,z\,\frac{
  \theta[\frac{M^2 y}{m^2}-\frac{M^2 y^2}{m^2}-y-z-\frac{1-y-z}{x-x^2}]}
  {M^2 y^2+(m^2-M^2)y+m^2 z} \nonumber \\
  +\hspace{-3mm}&&\frac{1}{512\pi^3 m^2}Im\int_0^1 d\,x\int_0^1 d\,y \Bigl{[}
  \ln^2(\frac{M^2 y^2}{m^2}-\frac{M^2 y}{m^2}+y-i\varepsilon)-
  \ln^2(\frac{M^2 y^2}{m^2}-\frac{M^2 y}{m^2}+1-i\varepsilon) \Bigr{]} \nonumber \\
  =\hspace{-3mm}&&\frac{\theta[M-m]\,\theta[2m-M]}{256\pi^3 m^2 M^2}
  (M^2-m^2)(\Delta-2\ln\frac{M^2-m^2}{m\,M}+2) \nonumber \\
  +\hspace{-3mm}&&\frac{\theta[M-2m]\,\theta[3m-M]}{256\pi^3 m^2 M^2}\Bigl{[}
  (M^2-m^2-M\sqrt{M^2-4m^2})\Delta+2(M^2-m^2)(1-\ln\frac{M^2-m^2}{m\,M}) \nonumber \\
  +\hspace{-3mm}&&M\sqrt{M^2-4m^2}(\ln\frac{M^2-4m^2}{m^2}+\frac{\pi}{\sqrt{3}}-4)
  \Bigr{]} \nonumber \\
  +\hspace{-3mm}&&\frac{\theta[M-3m]}{256\pi^3 m^2 M^2}\Bigl{[} (M^2-m^2-M\sqrt{M^2-4m^2})
  \Delta+M\sqrt{M^2-4m^2}(\ln\frac{M^2-4m^2}{m^2}+\frac{\pi}{\sqrt{3}}-4)
  \nonumber \\
  +\hspace{-3mm}&&2(M^2-m^2)(1-\ln\frac{M^2-m^2}{m\,M})
  +2M^2\int_{y_1}^{y_2}\left( F-\sqrt{3}\,tg^{-1}\frac{F}{\sqrt{3}}\right)d\,y \Bigr{]}\,,
\eeqa
where
\beqa
  F&=&\biggl{(}\frac{M^2 y^2-M^2 y-3m^2 y+4m^2}{y(M^2 y-M^2+m^2)}\biggr{)}^{1/2}
  \,, \nonumber \\
  y_{1,2}&=&\frac{M^2+3m^2\mp\sqrt{M^4-10m^2 M^2+9m^4}}{2M^2}\,.
\eeqa
Both Eq.(12) and Eq.(14) are too complex to be expressed by analytical formula, so we only compare their numerical results, as shown in Fig.10. From Fig.10 we find that the results obtained by the two methods coincide very well.
\setcounter{figure}{9}
\begin{figure}[htbp]
\begin{center}
  \epsfig{file=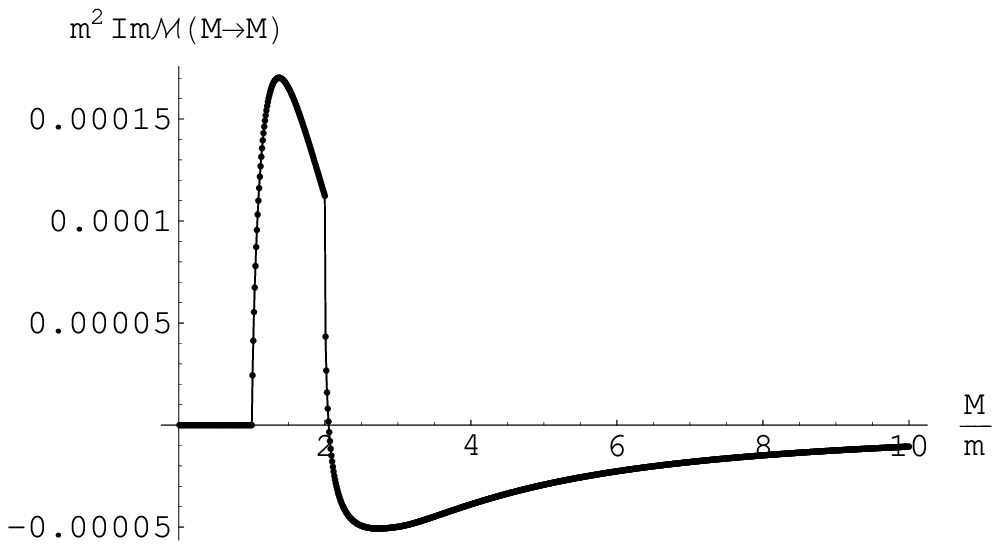,width=7.5cm} \\
  \caption{Numerical results of Eq.(12) (denoted by dot) and Eq.(14) (denoted by line) with $\Delta=0$.}
\end{center}
\end{figure}
In order to show to what extent the coincidence is we calculate the relative discrepancy of the two results, as shown in Fig.11.
\begin{figure}[htbp]
\begin{center}
  \epsfig{file=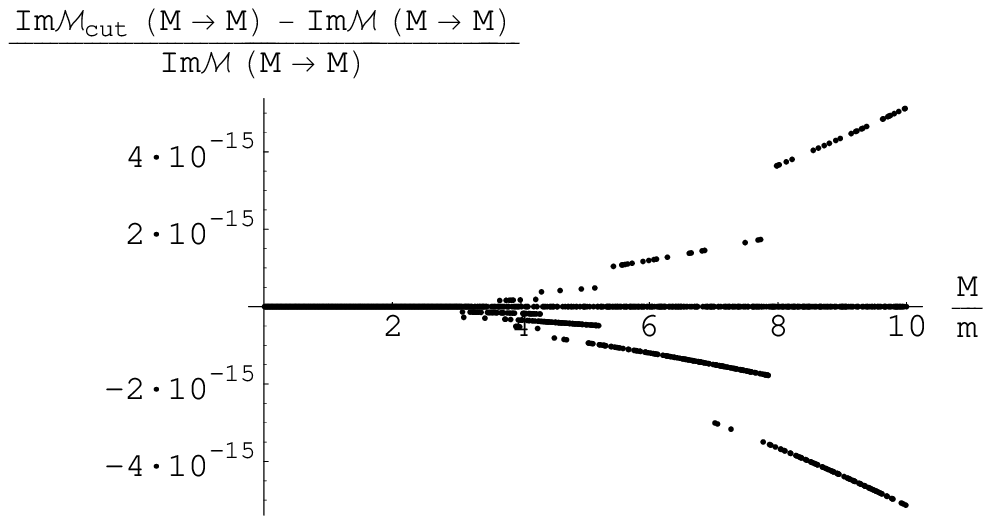, width=7.5cm} \\
  \caption{The relative discrepancy of Eq.(12) and Eq.(14). The amplitude with a subscript "cut" denotes
  it is obtained by the improved cutting rules, and the amplitude without subscript is obtained
  by the conventional integral algorithm.}
\end{center}
\end{figure}
Similarly as Fig.6 we have used the software {\em Mathematica 5.0} to calculate the result, but we have set the precision goal as $10^{-12}$. As we see from Fig.11, the maximum relative discrepancy of Eq.(12) and Eq.(14) is less than $10^{-14}$ which is less than the set precision goal, so we can draw the conclusion that Eq.(12) is equal to Eq.(14) at the precision $10^{-12}$. Besides, we note that a similar discussion has been present in Ref.\cite{c3}.

There are a lot of similar discussions which are less typical, so we don't list them here. All of the discussions show that the results of the improved cutting rules coincide with those of the conventional integral algorithm very well. So we think that the improved cutting rules is correct.

\section{Dynamic law of the intermediate on shell particles in quantum vacuum}

We have mentioned in section two that the improved cutting rules is based on the assumption that only the cut in which the
intermediate on shell particles propagate along a same direction--from the incoming particles to the outgoing particles--has contribution to Feynman amplitude. What can we get from the assumption? It is obvious that this assumption guarantees
two things: 1) the intermediate on shell particles cannot create themselves circularly in quantum vacuum, for example the cases of
the first and the third diagrams of Fig.1 are forbidden, thus the energy of any intermediate on shell particles is less than the
total energy of the incoming particles; 2) the time order of all the vertices connected by the intermediate on shell particles is
definite, since along the propagating direction of the intermediate on shell particles the time increases and the intermediate on shell particles propagate along a same direction--from the incoming particles to the outgoing particles (for example the time order of the vertices in Fig.3 is definite, and the case of indefinite time order of the vertices connected by intermediate on shell particles, like the cases of the first and the third diagrams of Fig.1, is forbidden), thus the time order of the processes of the intermediate on shell particles in quantum vacuum is definite, i.e. beginning from the incoming particles lastly ending at the outgoing particles.

On the other hand, from all of the cutting diagrams of this paper we find that a correct cutting must cut a Feynman diagram into several completely separated parts (this can be guaranteed by the constraint in the improved cutting rules that the least number of the cut propagators in any cut loop is two). Combining all of the above discussions we can get such a picture in mind about processes of the intermediate on shell particles in quantum vacuum: the incoming particles decay into intermediate particles, then the intermediate particles decay into the outgoing particles; or the incoming particles decay into intermediate particles and part of the outgoing particles, then the intermediate particles decay into the remaining outgoing particles; or part of the incoming particles decay into intermediate particles, then the intermediate particles and the remaining incoming particles decay into the outgoing particles; or the other intermediate processes at two level and beyond; all of the creation, propagation and annihilation of the intermediate particles are along a same direction: from the incoming particles to the outgoing particles, while the circular creation of the intermediate particles is forbidden. Thus we find that the processes of the intermediate on shell particles in quantum vacuum are same as real physical processes, the only difference between them is that the final states detected by experiments are the outgoing particles of the real physical processes, not the intermediate on shell particles, so we haven't noticed the existence of the processes of the intermediate on shell particles in quantum vacuum.

In fact in some special physical processes the process of the intermediate on shell particles can be observed. For example, when a physical colliding process happens through the way that the incoming particles fuse into an intermediate particle, then the intermediate particle decays into the outgoing particles, the curve of the cross section will present a peak at the position where the total energy of the incoming particles is equal to the mass of the intermediate particle \cite{c2}.  Obviously this peak in the curve of the cross section indicates the presence of the process of the intermediate on shell particle in quantum vacuum.

\section{Conclusion}

In this manuscript we discuss how to improve Cutkosky's cutting rules to make it more explicit and more practicable.  We point out that a propagator has two singularities, so the cutting rules must tell us which singularity is used in the calculations. In order to solve this problem we introduce a concept: {\em the propagating direction of the on shell propagator}, and we stipulate that the energy component of the momentum along the propagating direction of the on shell propagator must be positive. Thus the propagating direction of the on shell propagator represents the real propagating direction of the intermediate on shell particle in quantum vacuum. Under such stipulation which singularity being used will be given beforehand in a cut propagator with definite propagating direction. Through some careful analysis we acquire the improved cutting rules. In order to verify the correctness of the improved cutting rules we calculate the imaginary parts of several Feynman diagrams by two methods: the improved cutting rules and the conventional integral algorithm. We find that the results obtained by the two methods coincide very well. So the improved cutting rules should be correct.

Through the foundation process of the improved cutting rules we find that the processes of the intermediate on shell particles in quantum vacuum must obey the following rules: 1) they separate a Feynman diagram into several completely separated parts; 2) all of the creation, propagation and annihilation of the intermediate on shell particles are along a same direction--from the incoming particles to the outgoing particles; 3) the circular creation of the intermediate on shell particles (for example see the cases of the first and the third diagrams of Fig.1) is forbidden, and the energy of any intermediate on shell particles is less than the
total energy of the incoming particles. Thus the processes of the intermediate on shell particles in quantum vacuum are same as real physical processes, the only difference between them is that the final states detected by experiments are the outgoing particles of the real physical processes, not the intermediate on shell particles. In other words, only the process of the intermediate on shell particles that is same as real physical process can exist in quantum vacuum.

The present conclusion is only a pilot study. The problem of the intermediate on shell particles in quantum vacuum needs more investigations. We can expect that through more investigations people can reveal more and more secrets of the quantum field vacuum.

\vspace{5mm} {\bf \Large Acknowledgments} \vspace{2mm}

The author thanks Prof. Cai-dian Lu for the fruitful discussions and the corrections of the words.

\end{document}